\documentclass[10pt,twocolumn]{article}
\usepackage{times}
\usepackage[utf8x]{inputenc}
\usepackage[final]{pdfpages}
\usepackage{t1enc}
\usepackage{graphicx}
\usepackage{textcomp}
\usepackage[T1]{fontenc}

\newcommand{\Section}[1]{\vspace{-8pt}\section{\hskip -1em.~~#1}\vspace{-3pt}}

\begin{document}

\twocolumn[
\centering{\LARGE \bf Gaia-GOSA: An interactive service for coordination of asteroid observation campaigns} 

 \vskip10mm

 \begin{flushleft}
{\small {\bf T. Santana-Ros}, P. Bartczak and T. Micha{\l}owski\\
Astronomical Observatory Institute, Faculty of Physics, Adam Mickiewicz University, S{\l}oneczna 36, 60-286 Pozna{\'n}, Poland (tonsan@amu.edu.pl)\\} 
\end{flushleft}

\vskip10mm %
]

\thispagestyle{empty}

\section*{Abstract} 

We present the Gaia-Groundbased Observational Service for Asteroids (GOSA). Gaia-GOSA is an interactive tool which supports observers in planning photometric observations of asteroids. Each user is able to personalise the observation plan taking into account the equipment used and the observation site. The list of targets has been previously selected among the most relevant and scientifically remarkable objects, while the prediction of the transits in the Gaia's field of view have been calculated at the Observatoire de la C\^{o}te d'Azur. The data collected by the GOSA community will be exploited to enhance the reliability of the Gaia's Solar system science. The service is publicly available at www.gaiagosa.eu.

\Section{Introduction} 

To help amateur observers playing an important role in professional astronomy and provide
valuable observation record that will be used when analysing the Solar System Objects (SSO) Gaia
release, we have created a WWW service called Gaia-Groundbased Observation Service for
Asteroids (Gaia-GOSA). The service is publicly available at www.gaiagosa.eu. Any user of the service is
able to easily create an account, acquire the list of observation targets (in accordance with the
Gaia observations) adapted for the specific user's instrument and observation site. The users are also able to upload the gathered observation data to the server as the material for further research. These data will help the astronomers disentangling problematic inversion cases of the Gaia data, confirm suspected binaries or determine and improve asteroids’ synodic period measurements.
The service is available for the general public, but the registered users are able to enjoy a more personalised service experience: personalised observation planner, social networking with other observers, take part of the users ranking based on the scientific value of uploaded data, etc.

\Section{Transits of Solar system objects in 2015}

\begin{figure}[h]
\centerline{\includegraphics[width=\columnwidth]{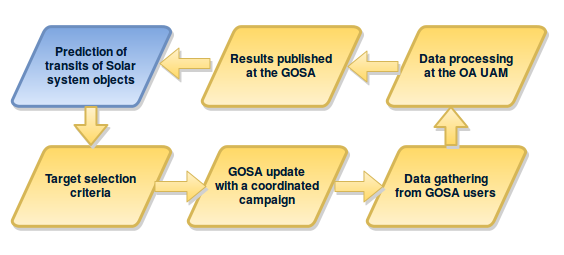}}
\caption{Flowchart of the GOSA data processing design.}
\label{flow}
\end{figure}

The Gaia-GOSA aims to release to the general public part of a highly specialized product generated within the
Gaia-DPAC Solar system group. In order to make this product useful to the general community it should
accomplish the following points:
\begin{enumerate}
\item It shall be understandable without any previous knowledge of the Gaia mission
\item It cannot contain specialized information or too many data
\item It shall provide clear instructions on how to use the product
\end{enumerate}
The prediction of transits of Solar system objects will be calculated approximately once per year at the Observatoire de la C\^{o}te d'Azur (OCA), due to the impossibility of predicting some of the input parameters for a longer period. These predictions were generated by F. Mignard from the OCA using the transit predictor developed within the DPAC-CU4/SSO \cite{mignard}. For each release, we aim to
select the most interesting targets under a scientific criteria and publish their predictions at the GOSA. In return, we expect gathering observations from the users, which will be analysed at the Astronomical Observatory of the Adam Mickiewicz University (AO AMU). The results obtained will be published at the GOSA site as well as in specialized journals, while feedback about the observation quality will be sent to the users. The full expected flowchart of the data processing design is shown in Fig.~\ref{flow}.

\Section{Combining Gaia observations with ground-based lightcurves} 

Combining Gaia observations of asteroids with ground-based lightcurves becomes straightforward when
both observations are taken simultaneously. In contrast, if the lightcurve obtained from the ground does not
include the epoch of observation by Gaia, there may be problems to link the Gaia observation to the
rotational phase, and to calibrate the magnitudes of the ground-based data, especially in cases when the
lightcurve is complex and the period resulting from the lightcurve is uncertain.
The GOSA is publishing the Gaia observation sequence for selected Solar system objects, allowing the
observers to obtain a lightcurve of a certain asteroid at the same time as Gaia is collecting a very precise
photometric measurement. Later on, it will be possible to calibrate the ground-based observation (even if it is
relative photometry) with the Gaia absolute magnitude, and proceed with the inversion process normally.
Formally, the only difference between data sources will appear during the preparation of the input file
containing the photometric error associated with each observational instrument and the position vectors of
the observer.

\Section{Target selection criteria} 

Based on the results obtained in Santana-Ros et al. 2015 \cite{toni}, we defined the first target selection
criteria: additional ground-based observations are sought for asteroids with low pole latitudes, suspected
nearly spherical shape and with less than 50 Gaia transits. However, we are not able to predict the
transits of Solar system objects until the end of the mission, as the predictions require for regular corrections
due to unpredictable parameters, such as micrometeorite impacts on the spacecraft surface. Thus, for such
group of objects, it will be necessary to perform data mining at the end of the Gaia mission in order to
enhance their inversion results.
In addition to these results, we recommend to include as GOSA targets asteroids which might be
scientifically interesting for the Solar system science community. Such objects include:
\begin{itemize}
\item Known or suspected binary asteroids
\item Suspected fast rotators (period shorter than 1 hour) for which Gaia could have problems for the
period determination
\item Interesting categories, including strange spectral types, asteroid families, slow rotators, etc.
\end{itemize}
Based on these considerations, the AO AMU has prepared a list of interesting targets, which are included in GOSA. However, we consider this list to be dynamical, in the sense that more targets can (or shall) be included. Some examples might be: newly discovered NEOs,
targets of opportunity for other observational techniques (radar echo, adaptive optics, stellar occultation), objects for which a new relevant discovery has been done.
For this reason, we allow GOSA users to submit requests to add objects to the target list if they can
show them to be scientifically meaningful.

\Section{Summary and Conclusions}

We have created an interactive tool which allows observers from all around the world to plan their observations with a concrete scientific goal (enhance the Gaia Solar system science), coordinate with other observers in observational campaigns, and feel the excitement of being part of a real space mission such as Gaia. We encourage all the observers to register for free at the GOSA site (www.gaiagosa.eu) and spread the word to other astronomy fans.

\section*{Acknowledgements}

We thank F. Mignard and P. Tanga (OCA, Nice) for putting at our disposal the transit predictions of Solar system objects. The web service has been developed by OA AMU in collaboration with ITTI Sp. z o.o. while the data processing is provided by OA AMU. The service has been funded under the ESA Contract No. 400011266014/NL/CBi: "Gaia-GOSA: An interactive service for asteroid follow-up observations.".

\end{document}